\documentclass[11pt]{article}
\usepackage{natbib,moriond}
\usepackage{amsmath}
\usepackage{amssymb}
\usepackage{xspace}
\usepackage{accents}

\usepackage{epstopdf}
\usepackage{thumbpdf}

\def\Journal#1#2#3#4{{#1} {\bf #2}, #3 (#4)}

\newcommand{\mathbfit}[1]{\textbf{\textit{#1}}}

\newcommand{\rmn}[1]{\mathrm{#1}}
\newcommand{\ltsima}{$\; \buildrel < \over \sim \;$}
\newcommand{\lsim}{\lower.5ex\hbox{\ltsima}}
\newcommand{\gtsima}{$\; \buildrel > \over \sim \;$}
\newcommand{\gsim}{\lower.5ex\hbox{\gtsima}}
\newcommand {\apgt} {\ {\raise-.5ex\hbox{$\buildrel>\over\sim$}}\ }
\newcommand {\aplt} {\ {\raise-.5ex\hbox{$\buildrel<\over\sim$}}\ } 

\newcommand{\bra}{\langle}
\newcommand{\ket}{\rangle}

\newcommand{\vel}{\upsilon}

\newcommand{\e}{\rmn{e}}
\newcommand{\p}{\rmn{p}}

\newcommand*{\pbar}[1]{\accentset{(-)}{#1}}

\def\s{{\rm s}} 
\def\Ms{{\rm M}\s} 
\def\yr{{\rm yr}} 

\def\Ms{M_\odot} 

\def\erg{{\rm erg}} 

\def\Fermi{{\em Fermi}\xspace}

\DeclareSymbolFont{UPM}{U}{eur}{m}{n}
\SetSymbolFont{UPM}{bold}{U}{eur}{b}{n}
\DeclareSymbolFont{AMSa}{U}{msa}{m}{n}
\DeclareMathSymbol{\upi}{0}{UPM}{"19}
\DeclareMathSymbol{\umu}{0}{UPM}{"16}
\DeclareMathSymbol{\upartial}{0}{UPM}{"40}

\SetSymbolFont{symbols}{bold}{OMS}{cmsy}{b}{n}
\DeclareSymbolFont{bmisymbols}{OML}{cmm}{b}{it}
\DeclareMathSymbol{\balpha}{0}{bmisymbols}{"0B}
\DeclareMathSymbol{\bbeta}{0}{bmisymbols}{"0C}
\DeclareMathSymbol{\bgamma}{0}{bmisymbols}{"0D}
\DeclareMathSymbol{\bdelta}{0}{bmisymbols}{"0E}
\DeclareMathSymbol{\bepsilon}{0}{bmisymbols}{"0F}
\DeclareMathSymbol{\bzeta}{0}{bmisymbols}{"10}
\DeclareMathSymbol{\boldeta}{0}{bmisymbols}{"11}
\DeclareMathSymbol{\btheta}{0}{bmisymbols}{"12}
\DeclareMathSymbol{\biota}{0}{bmisymbols}{"13}
\DeclareMathSymbol{\bkappa}{0}{bmisymbols}{"14}
\DeclareMathSymbol{\blambda}{0}{bmisymbols}{"15}
\DeclareMathSymbol{\bmu}{0}{bmisymbols}{"16}
\DeclareMathSymbol{\bnu}{0}{bmisymbols}{"17}
\DeclareMathSymbol{\bxi}{0}{bmisymbols}{"18}
\DeclareMathSymbol{\bpi}{0}{bmisymbols}{"19}
\DeclareMathSymbol{\brho}{0}{bmisymbols}{"1A}
\DeclareMathSymbol{\bsigma}{0}{bmisymbols}{"1B}
\DeclareMathSymbol{\btau}{0}{bmisymbols}{"1C}
\DeclareMathSymbol{\bupsilon}{0}{bmisymbols}{"1D}
\DeclareMathSymbol{\bphi}{0}{bmisymbols}{"1E}
\DeclareMathSymbol{\bchi}{0}{bmisymbols}{"1F}
\DeclareMathSymbol{\bpsi}{0}{bmisymbols}{"20}
\DeclareMathSymbol{\bomega}{0}{bmisymbols}{"21}
\DeclareMathSymbol{\bvarepsilon}{0}{bmisymbols}{"22}
\DeclareMathSymbol{\bvartheta}{0}{bmisymbols}{"23}
\DeclareMathSymbol{\bvarpi}{0}{bmisymbols}{"24}
\DeclareMathSymbol{\bvarrho}{0}{bmisymbols}{"25}
\DeclareMathSymbol{\bvarsigma}{0}{bmisymbols}{"26}
\DeclareMathSymbol{\bvarphi}{0}{bmisymbols}{"27}

\newcommand{\Photo}{}

\bibpunct{(}{)}{;}{a}{,}{,}

\begin{document}
\vspace*{4cm}
\title{INTRODUCTION TO EXTRAGALACTIC SOURCES OF VERY HIGH-ENERGY PHOTONS}

\author{ C.~PFROMMER }

\address{Heidelberg Institute for Theoretical Studies, Schloss-Wolfsbrunnenweg 35, 69118 Heidelberg, Germany}

\maketitle\abstracts{The launch of the \Fermi gamma-ray space telescope and the
  imaging air Cerenkov telescopes H.E.S.S., MAGIC, and VERITAS have
  substantially transformed our knowledge of gamma-ray sources in the last
  decade. The extragalactic gamma-ray sky is teeming with blazars, which are
  active galactic nuclei whose jet is directed at us. Additionally, there are
  radio galaxies, starburst and spiral galaxies, and gamma-ray bursts, albeit
  with smaller numbers. Galaxy clusters have not yet been observed in gamma
  rays. Here, I will introduce the different gamma-ray emission processes and
  review what they may tell us about these objects and the underlying
  acceleration mechanisms. Beyond the study of these fascinating objects, TeV
  gamma rays from blazars probe the integrated star formation history of the
  universe. Studies of TeV blazar spectra may provide us with insights into
  intergalactic magnetic fields or alternatively, may lead us to infer the
  existence of a novel mechanism that heats the intergalactic medium at late
  times (for redshifts $z<3$) and impacts the Lyman-$\alpha$ forest and
  late-time structure formation. The TeV gamma-ray emission may also allow us to
  probe fundamental physics such as the structure of space time.}

\section{Introduction}
\label{sec:intro}

Gamma rays cannot penetrate the Earth's atmosphere to the ground. To directly
detect them, we have to go to space, where the Large Area Telescope onboard
\Fermi is currently surveying the entire sky at energies ranging from around 100
MeV to 100 GeV. The upper energy limit is determined by the finite size of the
detector, which uses the pair conversion technique to detect gamma rays.
However, for a gamma-ray energy above 30 GeV, it is feasible to detect the flash
of Cerenkov light that is emitted in the electromagnetic cascade initiated by
the gamma ray penetrating the Earth's upper atmosphere. This allows to
reconstruct the energy and arrival direction of the original gamma ray and opens
a second window to the gamma-ray sky that is employed by the imaging air
Cerenkov telescope collaborations H.E.S.S., MAGIC, and VERITAS. Those have to
conduct pointed observations of interesting targets. While this allows deeper
observations on single objects with better angular resolution, it complicates
the characterization of the selection function for population studies.  What are
the main questions that we can address by observationally probing the
extragalactic gamma-ray sky?
\begin{itemize}
\item Which objects can we see? Clearly, we can do astronomy in the gamma-ray
  band and characterize and classify the detected objects into
  populations. Using multi-frequency constraints, we can then identify the
  objects with known counterparts at other electromagnetic
  wavelengths. Gamma-ray emitting objects are active galactic nuclei (blazars,
  radio galaxies), starburst and spiral galaxies, and gamma-ray bursts.
\item What underlying physics can we probe? In practice, gamma-ray emission
  probes the most extreme physics laboratories of the cosmos. This allows us to
  assess questions about the mechanism of particle acceleration and magnetic
  field amplification, and to study plasma physical processes at conditions that
  are quite different from those achievable in our laboratories on Earth (e.g.,
  collisionless plasmas, extreme Lorentz factors).
\item What fundamental physics can we hope to learn?  (1) TeV photons produce
  electron-positrons pairs upon annihilating with soft photons in the
  extragalactic background light. This process enables us to probe the
  integrated star formation history of the universe. (2) It also may provide us
  with insights into intergalactic magnetic fields (which could be of primordial
  origin). (3) If the kinetic pair energy can be efficiently dissipated, this
  may even provide the dominant energy source to the intergalactic medium at
  late times (for redshifts $z<3$) and hence impact the intergalactic medium,
  the Lyman-$\alpha$ forest and late-time structure formation.  (4) Extreme
  variability at the highest gamma-ray energies may enable us to probe the
  structure of space time.  (5) The gamma-ray energies are well adapted to the
  weak energy scale, which coincides with the dark matter particle masses in the
  most popular models since those weakly interacting massive particles would
  naturally account for the observed relic density if they had thermally
  decoupled in the early universe. Hence, studies of the physics associated with
  gamma rays can provide essential clues for our understanding of structure
  formation, cosmology, and particle physics.
\end{itemize}

After introducing the different gamma-ray emission processes, I will discuss
each object class and the physics that this allows us to probe. This is meant to
be a pedagogical introduction rather than a comprehensive review for which the
reader is referred to Rieger {\it et al.} (2013).

\section{Gamma-ray emission processes induced by cosmic rays}
\label{sec:emission}

Gamma rays carry complementary information to cosmic rays (CRs) and unlike the
latter, point back to their origin (except for the ultra-high-energy CRs that
perhaps are also little deflected by intervening magnetic fields). We
distinguish {\it hadronic processes} (as a result of CR proton- or ion-initiated
interactions) and {\it leptonic processes} due to relativistic electrons or
positrons. At the heart of all of these emission processes are the
acceleration mechanisms that produce a (non-thermal) population of relativistic
protons or electrons in first place. Gamma-ray observations can thus give us
some constraints on the underlying type of acceleration. \vspace{.5em}

\noindent
\begin{minipage}[h]{0.55\textwidth}
  {\em hadronic processes:}
  \begin{itemize}
  \item pion decay:
  \begin{equation}
    \nonumber
    \p + \p\,(\rmn{ion}) \to \left\{
      \begin{array}{lcl}
        \upi^0 & \to & \gamma\gamma \\
        \upi^\pm & \to & \e^\pm + \nu_\mu + \bar{\nu}_\mu + \pbar{\nu}_\e 
      \end{array}
    \right.
  \end{equation}\vspace{-1em}
\item photo-meson production:
  \begin{equation}
    \nonumber
    \p\,(\rmn{ion}) + \gamma \to  \left\{
      \begin{array}{lcl}
        \upi^0 & \to & \gamma\gamma \\
        \upi^\pm & \to & \e^\pm + \nu_\mu + \bar{\nu}_\mu + \pbar{\nu}_\e 
      \end{array}
    \right.
  \end{equation}\vspace{-1em}
\item Bethe-Heitler pair production:
  \begin{equation}
    \nonumber
    \p\,(\rmn{ion}) + \gamma \to  \p\,(\rmn{ion}) + \e^+ +  \e^-
  \end{equation}
  \end{itemize}
\end{minipage}
\hfill
\begin{minipage}[h]{0.44\textwidth}
  {\em leptonic processes:}
  \begin{itemize}
  \item inverse Compton:
  \begin{equation}
    \nonumber
    \e^* + \gamma \to \e + \gamma^*
  \end{equation}\vspace{-.25em}
  \item synchrotron radiation:
  \begin{equation}
    \nonumber
    \e^*\,(\p^*) + B \to \e\,(\p) + B + \gamma^*
  \end{equation}\vspace{0em}
  \item bremsstrahlung:
  \begin{equation}
    \nonumber
    \e^* + \p\,(\rmn{ion}) \to \e + \p\,(\rmn{ion}) + \gamma^*
  \end{equation}
  \end{itemize}
\end{minipage}\vspace{1em}

The hadronic p-p reaction proceeds if the center-of-momentum energy exceeds the
kinetic energy threshold of the pion rest mass (or more precisely the $\Delta$
resonance through which this reaction proceeds). The latter two {\it hadronic
  processes} require that the gamma-ray energy in the rest system of the proton
(ion) exceeds the rest mass of the pion or twice that of the electron,
respectively. This can either be achieved by a CR interacting with a soft photon
that is Lorentz boosted to the CR's rest frame or by an energetic gamma
ray. The latter could result from either of the {\em leptonic processes} (shown
on the right; a star denotes here an energetic particle).

Inverse Compton emission and synchrotron emission are very similar processes
since both describe the interaction of an energetic lepton with a photon. This
photon can either be provided by an astrophysical radiation field or is a
virtual photon that mediates the electromagnetic interaction of the electron
with the magnetic field $B$. Because the synchrotron emissivity of a particle
with mass $m$ scales with the Thompson cross section, $\sigma_T \propto m^{-2}$,
protons have to be more energetic by a factor of $(m_\p/m_\e)^2$ to produce an
emission that is comparable to that of electrons.  Finally, (non-)thermal
bremsstrahlung emission is caused by the acceleration of a (non-)thermal
electron in the Coulomb field of an ion. The gamma-ray spectra resulting from
these non-thermal processes are typically power-law spectra that reflect the
parent CR (electron or proton) power-law spectra and are convolved with the
respective emission spectrum of an individual particle. Changing to a
dimensionless integration variable determines the relation between spectral
indices of the CR and gamma-ray emission spectra.

\section{Active galactic nuclei}
\label{sec:AGN}

Active galactic nuclei (AGN) can launch relativistic jets that are powered by
accretion onto a central nucleus, presumably a supermassive black hole.  The
widely accepted AGN standard paradigm provides a unified picture of their
emission properties, which depend on the orientation of the AGN relative to the
line of sight (Urry {\it et al.}  1995). There exist two main classes of AGNs
that differ in their accretion mode and in the physical processes that dominate
the emission.
\begin{itemize}
\item {\em Thermal/disk-dominated AGNs.} Infalling matter assembles in a thin
  disk and radiates thermal emission with a range of temperatures. The
  distributed black-body emission is then Comptonized by a hot corona above the
  disk that produces power-law X-ray emission, which is a measure of the
  accretion power of the central object.  This class of objects are called QSOs
  or Seyfert galaxies and make up about 90\% of AGNs. They preferentially emit
  in the optical or X-rays and do not show significant nuclear radio
  emission. None of these sources have so far been unambiguously detected by
  \Fermi or imaging atmospheric Cherenkov telescopes because the Comptonizing
  electron population is not highly relativistic and emits isotropically,
  i.e. there is no relativistic beaming effect that boosts the emission.
\item {\em Non-thermal/jet-dominated AGNs.} Highly energetic electrons that have
  been accelerated in the relativistic jet interact with the jet magnetic field
  and emit synchrotron photons that range from the radio to X-ray. The same
  population of electrons can also Compton up-scatter any seed photon population
  either provided by the synchrotron emission itself ({\it synchrotron
    self-Compton scenario}) or from some other external radiation field such as
  ultraviolet (UV) radiation from the accretion disk or the infrared (IR)
  radiation from the surrounding torus ({\it external Compton scenario}). Hence
  the spectral energy distribution (SED) of these objects shows two distinct
  peaks. Alternatively, sufficiently energetic {\it proton synchrotron} photons
  can be converted into a pion in the Coulomb field of a proton. The neutral
  pion decays into gamma rays that trigger an electromagnetic cascade, which
  produces a spectrum of gamma rays ({\it proton-induced cascades}).  The
  luminosity of all these non-thermal emission components probes the jet power
  of these objects. Observationally, this leads to the class of radio-loud AGNs
  which can furthermore be subdivided into blazars (with the line of sight
  intersecting the jet opening angle) and non-aligned non-thermal dominated
  AGNs.
\end{itemize}
Blazars can further be subdivided into two main subclasses depending upon their
optical spectral properties: flat spectrum radio quasars (FSRQ) and BL Lacs.
FSRQs, defined by broad optical emission lines, have SEDs that peak at energies
below $1\,\rmn{eV}$, implying a maximum particle energy within the jet and
limiting the inverse-Compton scattered photons mostly to the soft gamma-ray
band.  It is presumably for this reason that no continuous TeV component has
been detected in an FSRQ (while their flare emission can sometimes reach TeV
energies).

\begin{figure}
\begin{minipage}{0.45\linewidth}
\centerline{\includegraphics[width=\linewidth]{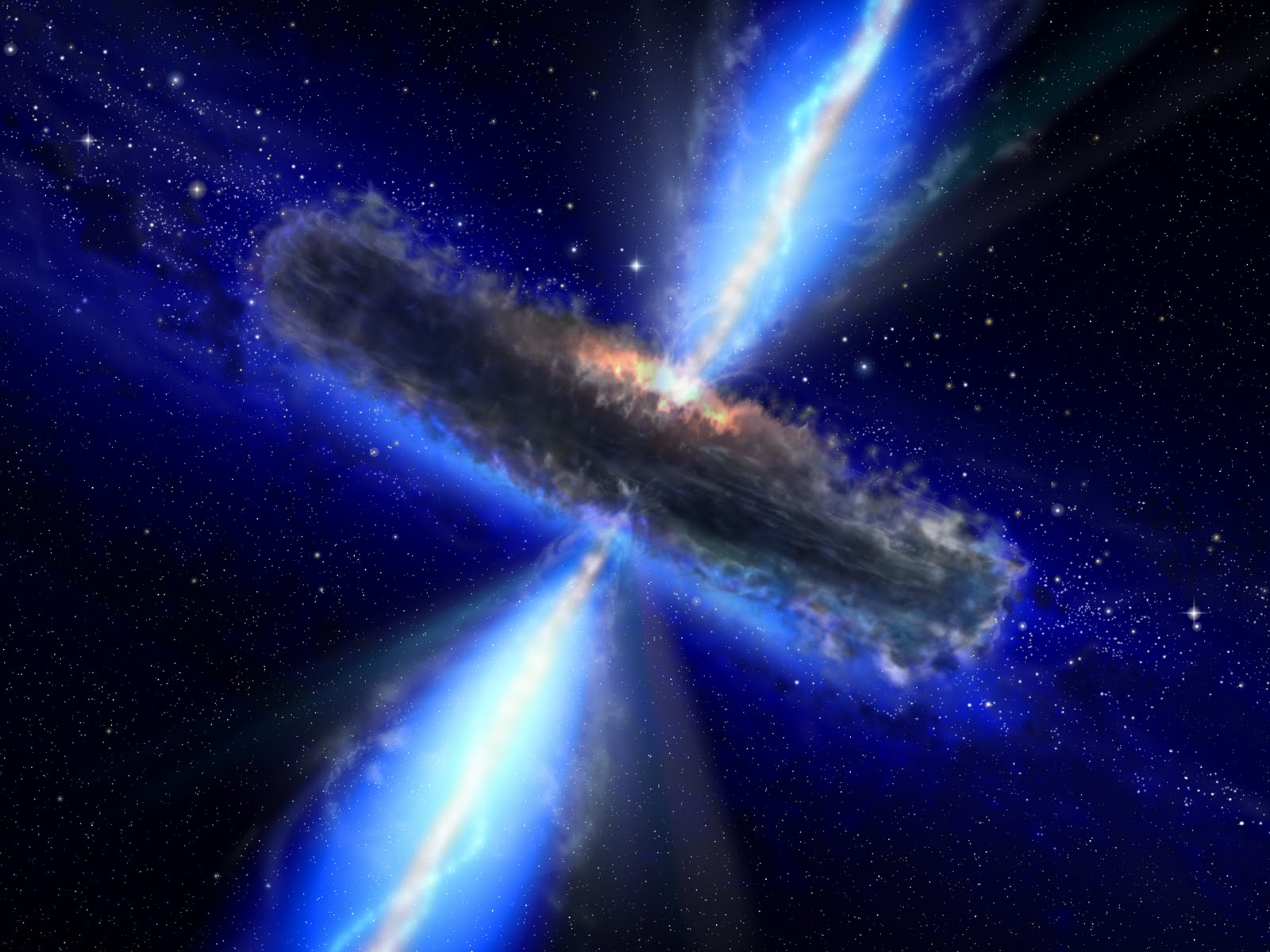}}
\end{minipage}
\hfill
\begin{minipage}{0.46\linewidth}
\vspace{2em}
\centerline{\includegraphics[width=\linewidth]{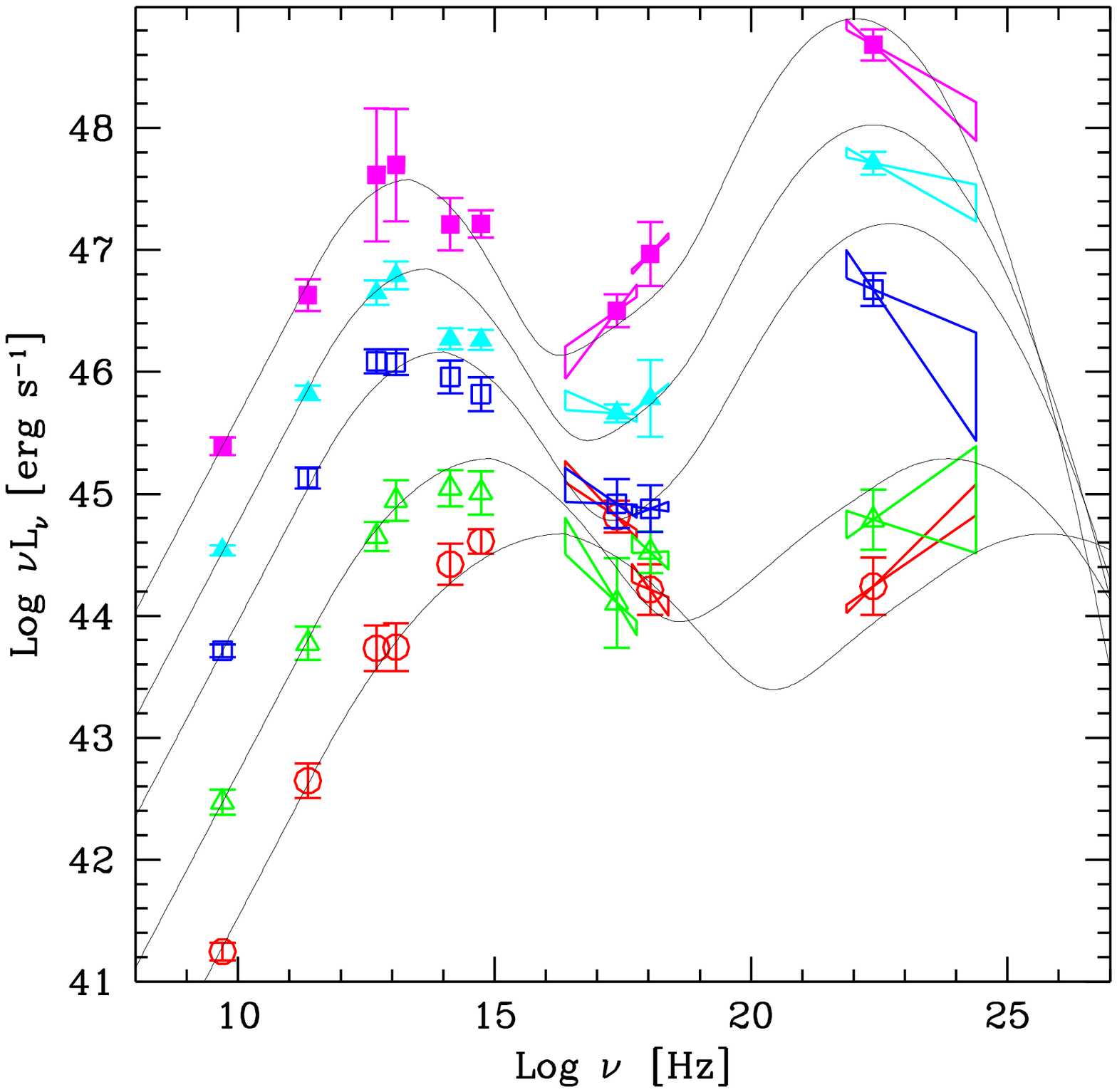}}
\end{minipage}
\vspace{-2em}
\caption{{\it Left.} Artist's impression of a super-massive black hole that
  launches relativistic jets and electromagnetic radiation out to cosmological
  distances. Blazars are systems where the jet opening angle contains our line
  of sight, i.e. the jet ``points'' at us (credit ESA/NASA/AVO/Padovani). {\it
    Right.}  Sequence of characteristic blazar SEDs as a function of source
  luminosity from FSRQ (top curve) to HSP objects (bottom curve, from Donato
  {\it et al.}  2001).}
\label{fig:blazar}
\end{figure}

In contrast, BL Lacs or Blazars of the BL Lac type (Massaro {\it et al.} 2009)
can be copious TeV emitters.  These are very compact radio sources and have a
broadband SED similar to that of strong lined blazars but lack the broad
emission lines that define those.  Depending upon the peak energy in the
synchrotron spectrum, which approximately reflects the maximum particle energy
within the jet, they are classified as low-, intermediate-, or high-synchrotron
peaked BL Lacs, respectively called LSP, ISP, and HSP.  While LSPs peak in the
far-IR or IR band, they exhibit a flat or inverted X-ray spectrum due to the
dominance of the inverse-Compton component (see Fig.~\ref{fig:blazar}). The
synchrotron component of ISPs peaks in the optical, which moves their
inverse-Compton peak into the gamma-ray band of \Fermi.  HSPs are much more
powerful particle accelerators, with the synchrotron peak reaching into the UV
or, in some cases, the soft X-ray bands. The inverse-Compton peak can then reach
TeV energies.

Hard \Fermi blazars (defined by a rising energy spectrum, $E^2 dN/dE$, in the
\Fermi band, i.e., HSPs and some ISPs) have a redshift distribution that is
peaked at low redshifts extending only up to $z=0.7$. This is most likely
entirely a flux selection effect; hard blazars are intrinsically less luminous
than LSPs and FSRQs, with an observed isotropic-equivalent luminosity range of
$10^{44} - 2\times10^{46}~\rmn{erg~s}^{-1}$, with the highest redshift hard
\Fermi blazars also being among the most luminous objects.  There are plausible
explanations why hard \Fermi blazars should be intrinsically less luminous than
FSRQs. Ghisellini {\it et al.} (2009) have argued that the physical distinction
between FSRQs and hard blazars has its origin in the different accretion regimes
of the two classes of objects. Using the gamma-ray luminosity as a proxy for the
bolometric luminosity, the boundary between the two subclasses of blazars can be
associated with the accretion rate threshold (nearly 1\% of the Eddington rate)
separating optically thick accretion disks with nearly Eddington accretion rates
from radiatively inefficient accretion flows.  The spectral separation in hard
(BL Lacs) and soft (FSRQs) objects then results from the different radiative
cooling suffered by the relativistic electrons in jets propagating into
different surrounding media (Ghisellini {\it et al.} 2009).  Hence in this
model, hard \Fermi blazars cannot reach higher luminosities than approximately
$2\times10^{46}\,\erg\,\s^{-1}$ since they are limited by the nature of
inefficient accretion flows that power these jets and by the maximum black hole
mass, $\sim10^{10}\,\Ms$.

Despite the tremendous progress in our understanding of properties of the
various populations, there are many open questions, including the jet
energetics, the mechanisms responsible for jet formation and collimation, the
plasma composition (leptonic vs. hadronic), the jet geometry (1-zone
vs. spine-layer), or the specific acceleration mechanisms of the jet
plasma. Particularly puzzling is the reason for the observed variability of the
gamma-ray emission of blazars, which is considerably smaller than the light
crossing time of the Schwarzschild horizon in the most extreme cases. TeV
``flares'' may sign instabilities in the accretion of matter onto the central
supermassive black hole or in the jet.

\section{The impact of TeV blazars on cosmology and structure formation}
\label{sec:blazars}

\begin{figure}
\hfill
\begin{minipage}{0.24\linewidth}
\centerline{\includegraphics[width=\linewidth]{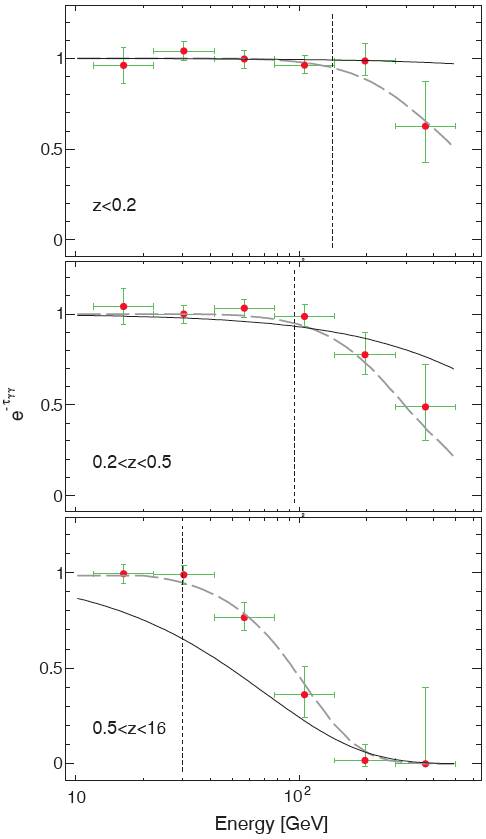}}
\end{minipage}
\hfill
\begin{minipage}{0.6\linewidth}
\vspace{1em}
\centerline{\includegraphics[width=\linewidth]{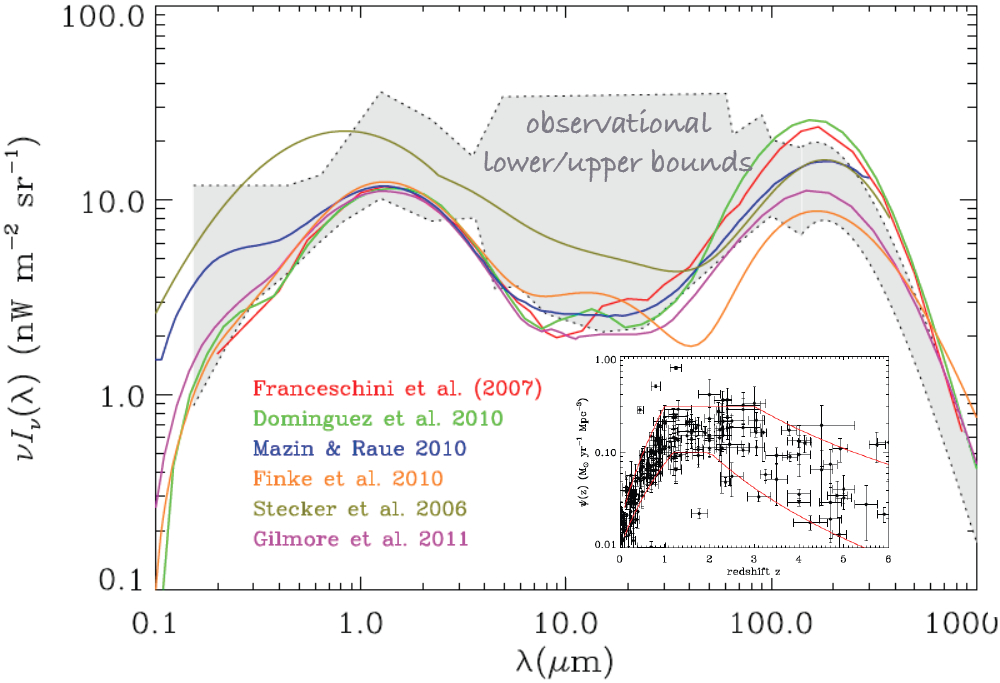}}
\end{minipage}
\hfill
\caption{{\it Left.} Stacked spectra of BL Lac objects show an absorption
  feature that moves to lower energies for increasing redshift (data points,
  from top to bottom). This confirms that gamma rays are attenuated by
  annihilating and pair producing on the EBL (dashed curve) and rules out models
  where all blazars have an intrinsic exponential cutoff and follow the blazar
  sequence (thin solid, from M. Ackermann {\it et al.}  2012). {\it Right.}
  Models of the EBL are compared to observational limits on the EBL. The inset
  shows a compilation of the cosmic star formation rate as inferred from UV,
  H$\alpha$, mid-IR, submillimeter, radio, and Ly$\alpha$ observations while
  excluding lower limits (from Dwek \& Krennrich 2012).  }
\label{fig:EBL}
\end{figure}

The light emitted by galaxies and accreting compact objects throughout the
history of the universe is encoded in the intensity of the {\it extragalactic
  background light} (EBL). Hence it provides an important integral constraint on
the star and quasar formation history in the hierarchical model of galaxy
assembly. Direct measurements of the EBL are limited by galactic and other
foreground emissions. Instead one can infer it indirectly because the universe
is opaque to TeV gamma rays, which annihilate and pair produce on the EBL. This
implies an absorption feature (the ``gamma-ray horizon'') in the spectra of
gamma-ray blazars. Stacking the spectra of 150 significantly detected BL Lac
blazars ($0.03<z<1.6$), the \Fermi Collaboration showed that the stacked
spectrum is unabsorbed for $E<25$~GeV. In agreement with the expectation, there
is an absorption feature that moves to lower gamma-ray energies for higher
source redshifts (propagation distances) due to attenuation of gamma rays by the
EBL at optical to UV frequencies (Ackermann {\it et al.} 2012, see
Fig.~\ref{fig:EBL}).

The ultra-relativistic pairs of electrons and positrons resulting from TeV
photon annihilation on the EBL are commonly assumed to lose energy primarily
through inverse Compton scattering with photons of the cosmic microwave
background, cascading the original TeV emission a factor of $\sim10^3$ down to
GeV energies. However, the expected cascaded GeV emission is not seen in the
individual spectra of those blazars (Neronov \& Vovk 2010). As a putative
solution to this problem, {\it intergalactic magnetic fields} have been
hypothesized, which would deflect the pairs out of our line-of-sight to these
blazars, diluting the point-source flux into a lower surface brightness ``pair
halo''.  A stronger magnetic field implies more deflection and dilution of the
GeV point source flux. In this picture, a non-detection of GeV gamma rays
suggests a limit on intergalactic magnetic fields of
$B\gtrsim(10^{-17}-10^{-15})\,\mu$G for a magnetic coherence length of 1 Mpc,
where the range covers uncertainties about the time delay of the cascade photons
(Taylor {\it et al.} 2011). Since most of the volume of the universe (and hence
a random sight line) is dominated by voids, magnetic fields of these strengths
may imply a primordial origin and allow one of these rare glimpses into the
early universe.

However, it has been shown recently that there is an even more efficient
mechanism that competes with this cascading process. The ultra-relativistic
pairs, originating from TeV photon annihilation on the EBL, are propagating
trough the intergalactic medium, which can be viewed as two counter-propagating
beams that are subject to plasma instabilities. The linear growth rate of the
so-called ``oblique instability'' is larger than the inverse Compton cooling
rate of the pairs. If this dominance of the instability growth rate carries over
to the regime of non-linear saturation, this implies a transfer of free kinetic
energy of the pairs to the unstable electromagnetic modes in the background
plasma, which should eventually be dissipated, heating the intergalactic medium
(Broderick {\it et al.} 2012, Schlickeiser {\it et al.} 2012). Typically,
$\sim300\,\yr$ after the onset of TeV emission, the pair beam density has grown
sufficiently for plasma beam instabilities to dominate its evolution, randomize
the beam, and potentially suppress the inverse-Compton signal upon which the
limits on the intergalactic magnetic fields are based (rendering these limits
dubious).  In this picture, there are two means to avoid the consequences of
plasma beam instabilities during the growth of the pair beam by (1) the sudden
appearance of a TeV-bright blazar or intrinsically transient sources (e.g.,
gamma-ray bursts) or (2) for particularly dim sources,
$L\lesssim10^{42}\,\erg\,\s^{-1}$, for which the pair beam density is too small
to support collective plasma behaviour. However, for all luminous TeV blazars
detected to date, the presence of these plasma beam instabilities appears
unavoidable and suggests the existence of a novel heating mechanism, coined {\it
  blazar heating}. It produces an inverted temperature-density relation of the
intergalactic medium (Chang {\it et al.}  2012) that is in agreement with
observations of the Lyman-$\alpha$ forest (Puchwein {\it et al.}  2012). This
also suggests that {\it blazar heating} can potentially explain the paucity of
dwarf galaxies in galactic halos and voids, and the bimodality of central gas
entropy values in galaxy clusters (Pfrommer {\it et al.} 2012).  Detailed
comparisons of predictions of blazar heating with \Fermi observation of blazar
statistics (redshift and $\log \mathcal{N}$-$\log S$ distribution) as well as
the isotropic and anisotropy gamma-ray backgrounds have been very successful and
supportive of this model (Broderick {\it et al.}  2013).

\section{Blazar and gamma-ray burst variability probes the structure of space time}
\label{sec:GRB}

Blazar variability shows a complex multi-wavelength behaviour that challenges
simple emission models. The H.E.S.S. observation of a giant flare (more than two
orders of magnitudes) of PKS 2155-304 shows a variability timescale $\Delta
t_\rmn{var} \sim (2-3)\,\rmn{min}\sim 0.02\, R_s/ c$, where $R_s/c$ is the light
crossing time of the Schwarzschild horizon (Aharonian {\it et al.}
2008). Causality requires $R<c \Delta t_\rmn{var} \gamma$ and implies a very
small emission region and bulk motion with a Lorentz factor $\gamma>50$
(Begelmann {\it et al.} 2008).

Independent of the emission mechanism, the observed variability can also be used
to probe the structure of space time and to constrain theories of Quantum
Gravity, some of which predict space-time to be ``foamy'' or discrete at the
Planck scale $l_P = \hbar/(m_P c)$, where the Planck mass is $m_P = \sqrt{\hbar
  c/G}$.  Preserving the $O(3)$ subgroup of $SO(3, 1)$, we can parametrize the
modified dispersion relation for photons, $c^2 \mathbfit{p}^2 = E^2 (1 + \xi
E/E_\rmn{QG} + \eta E^2/E_\rmn{QG}^2 + \ldots)$, where it is usually assumed
that $E_\rmn{QG}\sim m_P c^2 \sim 10^{19} \,\rmn{GeV}$ and $\xi=\pm1$ is a sign
ambiguity that is fixed in a given dynamical framework. Assuming that the
Hamiltonian equations of motion, $\dot{x}_i = \partial H/\partial p_i$, are
still valid, this yields $\vel=\partial E/\partial p = c\, (1 - \xi E/E_\rmn{QG}
+ \ldots)$. In other words, we obtain an energy-dependent time delay $\Delta t =
\xi (E/E_\rmn{QG}) (L/c) = 10 \,\rmn{ms}\,(\rmn{GeV}/E_\rmn{QG}) (1
\,\rmn{Gpc}/c) $, where $L$ is the path length of the photon. That is, we can
test this by studying propagation of high energy gamma-ray pulses of different
energies from cosmological distances (Amelino-Camelia {\it et al.}  1998).

That test was done with the \Fermi detection of the early arrival time of the 31
GeV photon of the short gamma-ray burst GRB 090510, implying a conservative
bound of $E_\rmn{QG} > 1.2 \times 10^{19}\,\rmn{GeV}$ (Abdo {\it et al.}
2009a). The giant flare of PKS 2155-304 observed by H.E.S.S. shows no observable
time delay between low- and high-energy photons, thereby implying a bound
$E_\rmn{QG} > 2.1 \times 10^{18}\,\rmn{GeV}$ (Abramowski {\it et al.}
2011). This starts to constrain an energy-dependent violation of Lorentz
invariance (i.e., an energy-dependent speed of light), which is predicted in
various models of Quantum Gravity. However, all these analyses make the strong
assumption that there is no intrinsic gamma-ray dispersion in the source and
that the gamma-ray pulses at different energies are emitted at the same time. To
improve upon these constraints, we need a better understanding of the sources
and emission mechanisms.

\section{Radio galaxies and the cluster ``cooling flow problem''}
\label{sec:RG}

\begin{figure}
\begin{minipage}{0.33\linewidth}
\centerline{\includegraphics[width=\linewidth]{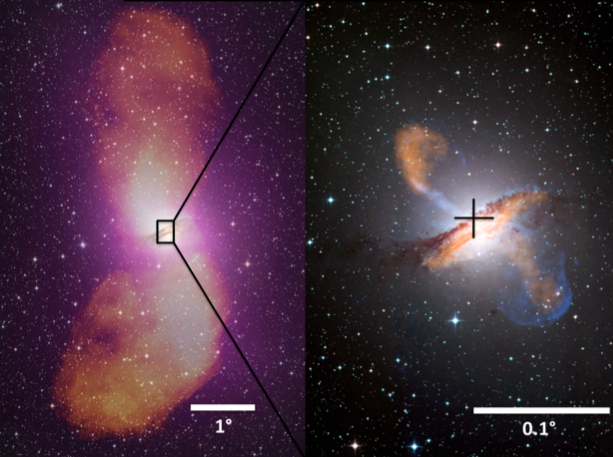}}
\end{minipage}
\hfill
\begin{minipage}{0.245\linewidth}
\centerline{\includegraphics[width=\linewidth]{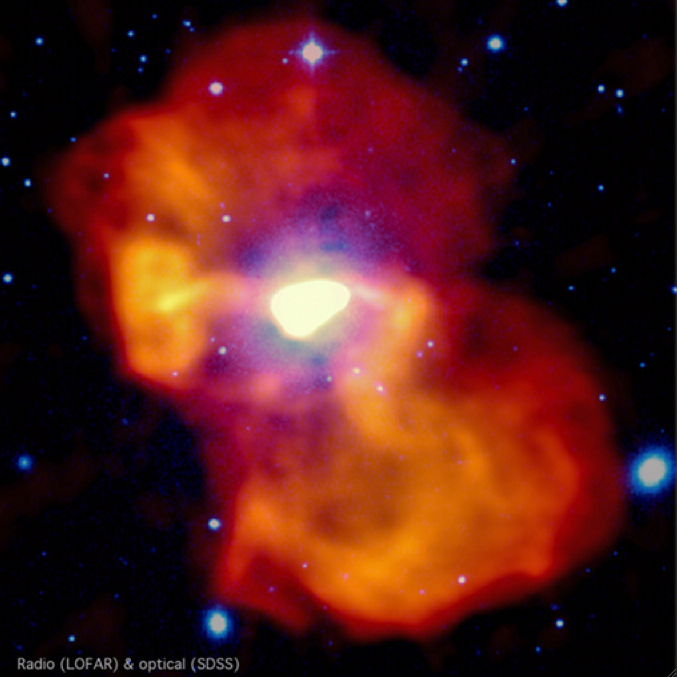}}
\end{minipage}
\hfill
\begin{minipage}{0.375\linewidth}
\vspace{0.75em}
\centerline{\includegraphics[width=\linewidth]{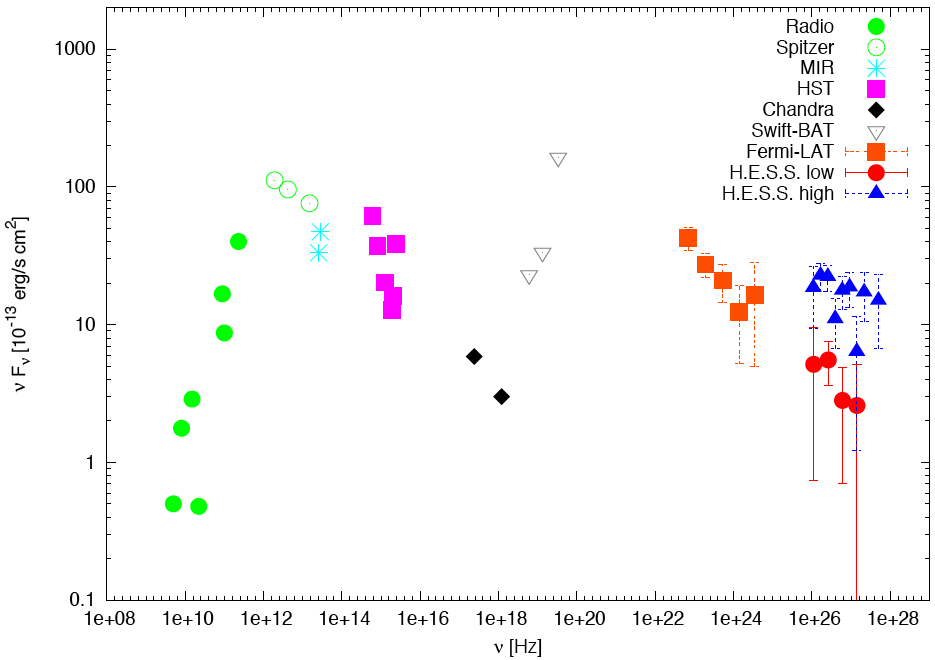}}
\end{minipage}
\caption{Nearby radio galaxies are emitting in gamma rays. \Fermi and
  H.E.S.S. detected gamma-ray emission from Cen A ({\it left}, radio data in
  orange are overlaid on \Fermi data in purple, credit NASA/DOE/\Fermi LAT) and
  M87 ({\it middle}, 140 MHz radio data from LOFAR/de Gasperin {\it et al.}
  2012). {\it Right.}  Multi-frequency spectrum of M87 indicating the low- (red)
  and high-flux state (blue) at TeV energies (from Rieger \& Aharonian 2012).}
\label{fig:RGal}
\end{figure}

Some nearby radio galaxies also emit gamma rays, which may be partly related to
the nucleus, the jet, the radio lobes, or CR interactions with the surrounding
plasma (see Fig.~\ref{fig:RGal}).  The closest radio galaxy Centaurus A is at a
distance of 3.7 Mpc, often considered as an ``AGN under the microscope''. \Fermi
observes GeV emission from the giant radio lobes and the core while
H.E.S.S. detected TeV emission from the nucleus/inner jet (Abdo {\it et al.}
2010, Aharonian {\it et al.} 2009). This triggered ideas that the giant lobes
are the sites of high-energy particle acceleration and production of
ultra-high-energy cosmic rays (Hardcastle {\it et al.} 2009).

At the end of the momentum-driven phase, relativistic jet particles inflate
radio-emitting lobes and do pressure-volume work on the ambient intra-group and
-cluster medium. According to the current paradigm, the buoyantly rising lobes
either do mechanical work on the surroundings (which gets dissipated through
shocks or a turbulent cascade) or they release CRs into the intracluster
medium. Those stream at the Alfv{\'e}n velocity with respect to the plasma rest
frame and heat the surrounding thermal plasma (Loewenstein {\it et al.}
1991). This ``AGN feedback'' balances radiative cooling and solves the cluster
``cooling flow problem'' at low redshifts, $z\lesssim 1$ (McNamara \& Nulsen
2012). What can gamma-ray observations add to this picture?

\Fermi and H.E.S.S. discovered gamma-ray emission from the radio galaxy M87
(Abdo {\it et al.} 2009b, Abramowski {\it et al.} 2012), the central galaxy of
the Virgo cluster, our closest galaxy cluster at a distance of 17 Mpc. While the
TeV emission in the high state is likely connected to the emission from the
nucleus/jet, there is the possibility that the low emission state traces
pion-decay gamma rays from the Virgo cool-core region as implied by the spectral
similarity to LOFAR radio data (see Fig.~\ref{fig:RGal}). In this picture, the
gamma-ray emission can be used to normalize the CR-induced heating rate, which
balances that of radiative cooling {\em on average} at each radius, thereby
suggesting a solution to the ``cooling flow problem'' in the Virgo cluster
(Pfrommer 2013).  This model would predict the gamma-ray emission in the low
state to be steady and slightly extended, which is testable with current
observations.

\section{Starburst and spiral galaxies}
\label{sec:SBG}

\begin{figure}
\centering{\includegraphics[width=0.6\linewidth]{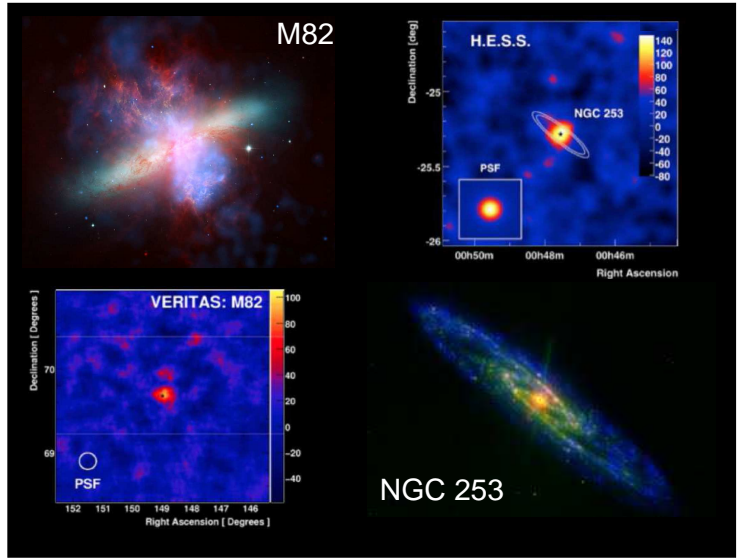}}
\caption{The nearby starburst galaxies M82 (Acciari {\it et al.} 2009) and NGC
  253 (Acero {\it et al.} 2009) emit TeV gamma rays. We contrast gamma-ray
  significance maps to optical images of these galaxies.}
\label{fig:starburst}
\end{figure}

M82 and NGC 253 are TeV gamma-ray emitting starburst galaxies (Acciari {\it et
  al.} 2009, Acero {\it et al.} 2009), both at a distance of $\sim3$ Mpc (see
Fig.~\ref{fig:starburst}). \Fermi confirmed the gamma-ray emission in those
objects and increased the sample of starburst galaxies by two more objects (NGC
4945, NGC 1068) although those are composite starburst/Seyfert 2 systems, which
makes it challenging to disentangle the pure starburst component (Lenain {\it et
  al.} 2010). Their star formation rate (in a compact region) is larger than
that of the Milky Way. In the emerging picture, supernova remnants, associated
with star formation regions, can energize CR protons through diffusive shock
acceleration.  Hadronic interactions of those CR protons with the ambient dense
gas produce pion-decay gamma rays. In the starburst region, there is dense
interstellar gas, with $\bra n\ket\sim250~\rmn{cm}^{-3}$, which yields a
hadronic interaction time that is of order the diffusive escape time, $t_{\p\p}
\sim t_\rmn{esc}$. Hence we are approaching the calorimetric limit.

The large magnetic field strengths and high densities should also give rise to
efficient leptonic emission. In fact, the tight far infrared (FIR)--radio
correlation implies universal conversion of the star formation rate to the CR-
and the synchrotron luminosities. Provided the picture of gamma-ray emission is
correct, this also would imply a FIR--gamma-ray correlation. The local four
spiral galaxies (Milky Way, SMC, LMC, M31) show indeed gamma-ray luminosities,
which fall on the locus of the FIR--gamma-ray correlation defined by the
starburst galaxies. However, to fully establish this picture, the AGN
contribution to the observed gamma-ray emission of starburst galaxies needs to
be carefully quantified. Moreover, the possible counterexamples to this
relation, i.e., upper limits on the gamma-ray emission of some galaxies that
fall also within the implied relation (IC342, NGC 6946), need to be understood.
While their upper limits are still compatible with the scatter around the
relation, tighter limits will either strengthen the tension or lead to a
detection, thereby confirming the existence of a FIR--gamma-ray correlation.

\section{Galaxy clusters}
\label{sec:clusters}

\begin{figure}
\begin{minipage}{0.36\linewidth}
\centerline{\includegraphics[width=\linewidth]{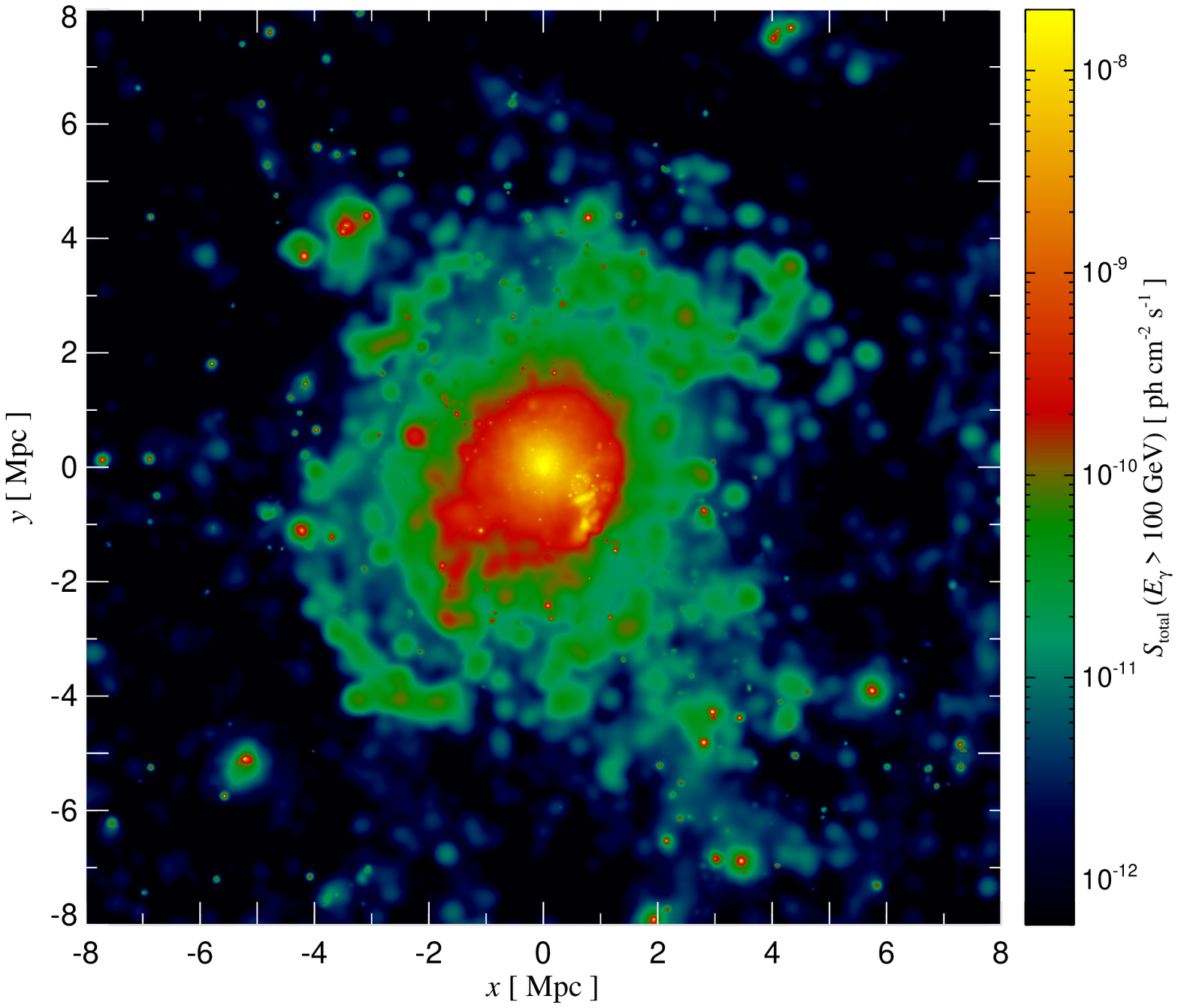}}
\end{minipage}
\hfill
\begin{minipage}{0.36\linewidth}
\centerline{\includegraphics[width=\linewidth]{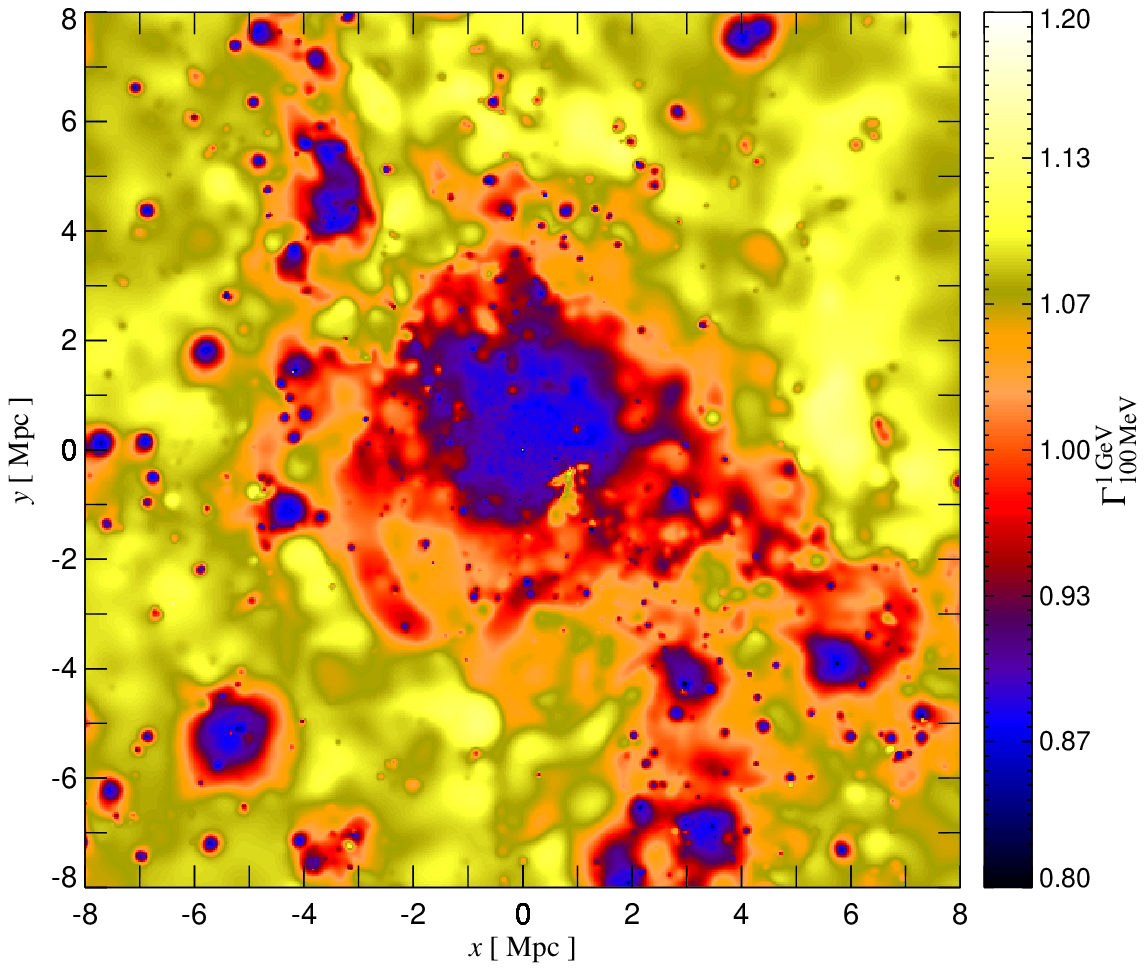}}
\end{minipage}
\hfill
\begin{minipage}{0.265\linewidth}
\centerline{\includegraphics[width=\linewidth]{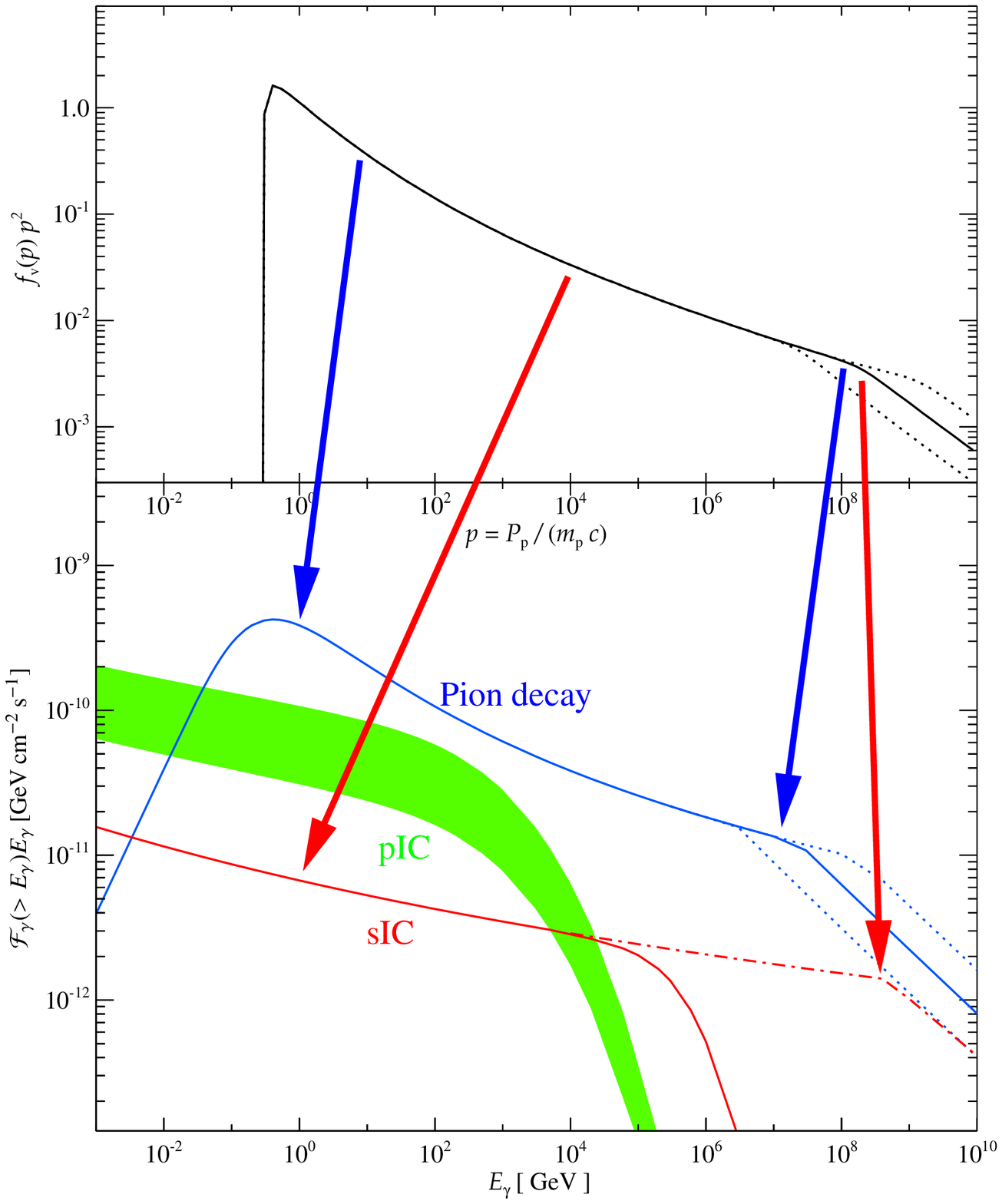}}
\end{minipage}
\caption{Expected gamma-ray emission in a cosmological hydrodynamical simulation
  of a galaxy cluster. The surface brightness emission ({\it left}) and spectral
  index map, $\Gamma_{100\,\rmn{MeV}}^{1\,\rmn{GeV}}$, ({\it middle}) is
  dominated in the center by pion decay and outside the virial radius (of 2.4
  Mpc) by primary inverse Compton (pIC) emission from shock-accelerated
  relativistic electrons. The total CR proton spectrum shows a universal shape
  across clusters ({\it top right}) that is inherited by the pion decay and
  secondary IC emission following hadronic p-p interactions ({\it bottom
    right}). The pIC emission only has a small contribution to the total
  emission for $E>30\,\rmn{MeV}$ (from Pinzke \& Pfrommer 2010).}
\label{fig:cluster}
\end{figure}

Despite many efforts in the recent years, no cluster-wide gamma-ray emission has
been detected so far (Ackermann {\it et al.} 2010, 2013). The observed radio halo and
relic emission on cluster scales (1-3 Mpc) proves the existence of CR electrons
and magnetic fields that permeate the cluster volumes and suggests that clusters
also emit diffuse gamma rays.  Cosmological hydrodynamical simulations of galaxy
cluster with self-consistent CR physics show that the normalized CR spectrum has
a universal concave shape across clusters (Pinzke \& Pfrommer 2010). During the
hierarchical assembly, every fluid element experienced on average the same
history of shock strengths, which is responsible for shaping the CR spectrum. As
a result, the gamma-ray signal is expected to be dominated by pion decay, but
other possibilities exist, such as inverse Compton emission from electrons that
have been accelerated at structure formation shock waves (see
Fig.~\ref{fig:cluster}).

Non-observations of gamma rays from the Perseus and Coma clusters constrain the
CR-to-thermal pressure to $P_\rmn{CR}/P_\rmn{th} < 1.7\%$ in those clusters
(Aleksic {\it et al.} 2012, Arlen {\it et al.} 2012). This immediately implies
that hydrostatic cluster masses are not significantly biased by CRs---an
important result if cluster populations are to be used for determining
cosmological parameters. A comparison to hydrodynamical cluster simulations
constrains the maximum acceleration efficiency at formation shock on average to
$<50\%$.  Provided the ({\it high-frequency}) radio halo emission is produced by
secondary electrons from CRp-p interactions, this allows us to place limits on
the central cluster magnetic fields of $>(4-9)\,\mu$G (Perseus) and
$>(2-5)\,\mu$G (Coma), which are below the limits obtained from Faraday rotation
measure studies (Aleksic {\it et al.}  2012, Arlen {\it et al.} 2012). However,
these limits on magnetic fields are in conflict with Faraday rotation data for
the {\it low-frequency} radio halo emission in Coma, arguing for a leptonic
origin of (at least) the external halo at these frequencies (Brunetti {\it et
  al.} 2011).

\section{Conclusions}
\label{sec:conclusions}

The non-thermal universe revealed by high-energy radiation provides not only
deep insights into high-energy astrophysics and plasma processes that are not
accessible in our laboratories on Earth but also new probes of
fundamental physics, cosmology, and structure formation. With the successful
\Fermi telescope and the imaging air Cerenkov collaborations H.E.S.S., MAGIC,
and VERITAS, we are currently entering a fascinating era that is complemented by
multi-frequency experiments. As a result of this, there is no shortage of new
discoveries and puzzles to solve. In tackling those, we should not be afraid of
employing new ideas and theories, some of which may later need to be refined. In
doing so, we should mind the unseen (dark matter, galaxy clusters, \ldots). What
can it teach us? Does it conflict with our previous models of particle
acceleration and/or transport?  To proceed, a precise and accurate measurement
of the isotropic and anisotropic extragalactic gamma-ray background is critical
for constraining the luminosity evolution of various populations by their
maximally allowed contribution to the respective backgrounds. I will end this
introduction by a quote of Louis Pasteur stating that ``in the fields of
observation chance favors only the prepared mind''!

\section*{Acknowledgments}

C.P.~gratefully acknowledges financial support of the Klaus Tschira Foundation
and thanks Chris Hayward for carefully reading the manuscript.

\end{document}